\documentclass[aps,prl,superscriptaddress,twocolumn,showpacs]{revtex4}
\usepackage{graphicx}

\begin{document}
\title{Whispering-gallery modes and light emission from a Si-nanocrystal-based single microdisk resonator}

\author{Mher Ghulinyan} 
\affiliation{Micro-Technology Laboratory, Fondazione Bruno Kessler,
Trento, I-38050 Trento, Italy}
\author{Daniel Navarro-Urrios}
\affiliation{Nanoscience Laboratory, Dept. Physics, University of
Trento, Via Sommarive 14, Povo, I-38050 Trento, Italy}
\author{Alessandro Pitanti}
\affiliation{Nanoscience Laboratory, Dept. Physics, University of
Trento, Via Sommarive 14, Povo, I-38050 Trento, Italy}
\author{Alberto Lui}
\affiliation{Materials and Analytical Methods Laboratory for
Biosensors and Bioelectronics, Fondazione Bruno Kessler, Trento,
I-38050 Trento, Italy}
\author{Georg Pucker}
\affiliation{Micro-Technology Laboratory, Fondazione Bruno Kessler,
Trento, I-38050 Trento, Italy}
\author{Lorenzo Pavesi}
\affiliation{Nanoscience Laboratory, Dept. Physics, University of
Trento, Via Sommarive 14, Povo, I-38050 Trento, Italy}

\begin{abstract}
We report on visible light emission from Si-nanocrystal based
optically active microdisk resonators. The room temperature
photoluminescence (PL) from single microdisks shows the
characteristic modal structure of whispering-gallery modes. The
emission is both TE and TM-polarized in $300$~nm thick microdisks,
while thinner ones (135~nm) support only TE-like modes. Thinner
disks have the advantage to filter out higher order radial mode
families, allowing for measuring only the most intense first order
modal structure. We reveal subnanometer linewidths and corresponding
quality factors as high as 2800, limited by the spectral resolution
of the experimental setup. Moreover,we observe a modification of
mode linewidth by a factor 13 as a function of pump power. The
origin of this effect is attributed to an excited carrier absorption
loss mechanism.
\end{abstract}


\maketitle
\section{Introduction}
High quality monolithic resonators such as micro-disks, rings and
toroids are triggering an intensive and rapidly evolving research.
Such structures, in which the total internal reflection leads to
circularly propagating optical modes, called whispering-gallery
modes (WGM) \cite{rayleigh}, are extremely attractive both from
device application and fundamental points of view \cite{vah}.
Optically passive microdisks, based on transparent materials with
negligible absorption losses, lead to extremely high quality factors
($Q\sim10^6$ to $10^{10}$), offering applications in spectroscopy
and sensing \cite{ilch_review,deio}. On the contrary, optically
active resonator systems, such as III-V semiconductor quantum dot
microdisk lasers, report active Q's of $10^3$--$10^{4}$ in the
visible and near infrared wavelength range \cite{lasers,kartik}.

The recent challenges in silicon photonics towards using
nanocrystalline Si (nc-Si) as an integrated light source have
boosted an intensive research in the last decade
\cite{pavesi,twdsilaser,sidev}. However, as an important cavity
system, nc-Si-based microdisk structures have been little studied
and only few works appear in the literature \cite{zacc,spie}. In
Ref.\cite{zacc} quality factors of few hundreds have been reported
for microdisk arrays of nc-Si/SiO$_2$ superlattices.

In this Letter we study the WGM emission properties of \emph{single}
microdisk resonators with an optically active disk material made of
luminescent nc-Si embedded in SiO$_x$ matrix. We report on
subnanometer WGM resonances corresponding to quality factors as high
as $2800$ around the wavelength of 800~nm, which to our knowledge
are the highest among the previously reported values in nc-Si-based
systems. We demonstrate the importance of exciting a single
resonator out of the mass-produced microdisk array in order to
reveal the fine modal structure in the light emission. Additionally,
we show an almost 13-fold narrowing of characteristic linewidths at
lowest excitation power associated to an attenuation of excited
carrier absorption losses.

\section{Sample preparation and micro-PL setup}
Our samples have been produced using standard silicon
microfabrication technology. The disk material has been realized
through plasma enhanced chemical vapor deposition (PECVD) of  135~nm
of Si rich silicon oxide (SRO) on top of crystalline silicon wafers.
A successive one hour annealing in an $N_2$ atmosphere at
1100$^{\circ}C$ leads to the formation of Si nanocrystals in the
SiO$_x$ host (with $\sim10\%$ of Si atoms in the nanocrystaline
phase). Then the wafers were photolithographically patterned and dry
etched anisotropically to form arrays of microdisk structures with
diameters ranging from 2$\mu$m to 10$\mu$m. Isolation of the
microdisks from the substrate was realized by an isotropic wet etch
of the latter, which formed mushroom-like devices.

\begin{figure}
\centering\includegraphics[width=6.5cm]{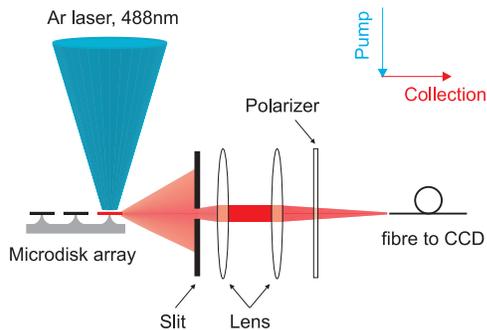} \caption{
(color online) The micro-PL setup; an individual microdisk is
excited vertically through a focused pumped beam, while the
characteristic WGM emission is collected under small solid angle in
the plane of the disks.} \label{setup}
\end{figure}

Room temperature micro-PL measurements have been performed using the
488~nm line of an Argon laser (see Fig.\ref{setup}). A long working
distance objective was used to focus the laser beam on the samples.
The microdisks close to the cleaved sample edge were excited
vertically, while the WGM emission was monitored in the plane of
disks, using collection optics (numerical aperture NA$=0.13$). A
polarizer was placed in the collection line to select transverse
electrically (TE) or transverse magnetically (TM) polarized PL
emission. Finally, the collected signal was sent to a spectrometer
interfaced to a cooled silicon charge coupled device (CCD). The PL
characterization was performed for microdisk arrays of various
diameters, while for simplicity we focus our attention on the
discussion of results obtained for the $8\mu$m diameter resonators.

\section{Whispering-gallery modes of single resonators}
When hundreds of resonators are excited simultaneously through a
large spot over the microdisk array, the observed emission lineshape
is characterized by roughly 5~nm wide WGM resonances (quality
factors of $\sim$160). Such low values can be attributed to an
inhomogeneous broadening of peaks due to slight dispersion of
microdisk diameters within the excited area \cite{zacc,note2}.

\begin{figure}  [t]
\centering\includegraphics[width=6.5 cm]{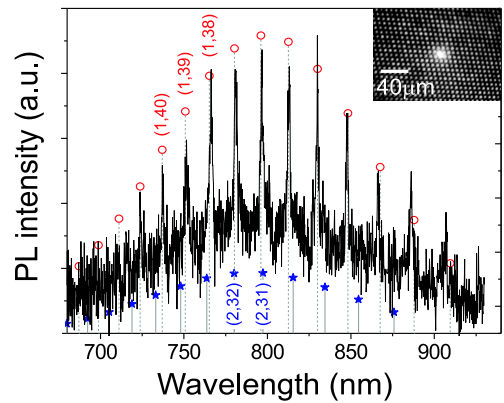} \caption{
(color online) Measured TE-polarized WGM spectrum of an 8$\mu$m
diameter microdisk is plotted together with the simulated peak
positions for the first radial mode family ($\circ$). The predicted
second mode family ($\star$) is however absent in the measured
spectrum because of being much less intense. (Inset) The bright spot
in the photograph is the direct image of the visible PL emission of
nc-Si from a single disk resonator. } \label{foto}
\end{figure}

To demonstrate this, we focused the excitation onto an individual
microdisk and recorded the resulting PL emission. In fact, from
Fig.\ref{foto} one observes immediately the fine WGM structure of
the single microdisk. The significant narrowing of the emission
lines results in subnanometer full width at half maxima (FWHM),
leading to quality factors of almost $3\times10^3$, limited by the
spectral resolution of our micro-PL setup (higher precision
measurements may reveal higher $Q$ values).

To get some insight on the light confinement inside the microdisk
resonator, we have performed effective index mode simulations of a
slab waveguide structure with the same thickness ($d$=135~nm) and
material refractive index (1.8 at 800~nm). For such choice of
parameters, we have obtained a well confined TE mode
($n_{eff}$=1.34) while an almost-radiative TM mode ($n_{eff}$=1.08).
Since these indices represent the upper limits for the
$n_{eff}$(TE,TM) of the 3D resonator structure, we therefore expect
that our microdisk does not support a guided TM mode. This is
confirmed both by experimental results and by 3D finite-difference
time-domain simulations using a freely available software package
\cite{ftdt}.

\begin{figure}  [b]
\centering\includegraphics[width=8 cm]{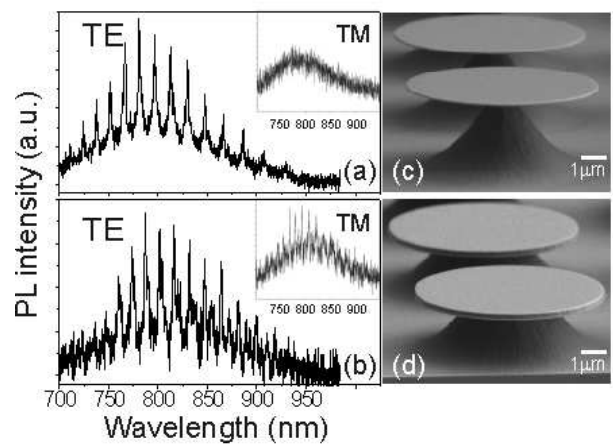} \caption{(color
online) Typical TE-polarized WGM emission spectra from $8\mu$m
diameter (a) thin, $d$=135~nm and (b) thick, $d$=300~nm disk
resonators. The corresponding insets show (a) the absence and (b)
the presence of resonant whispering-gallery features for
TM-polarization. Scanning electron microscopy images are shown
respectively in panels (c) and (d). Note, that though the support
pedestals have different top radii, in both cases they are small
enough to not disturb the WGM characteristics. } \label{tmte}
\end{figure}

Indeed, as one can see from the inset of Figure \ref{tmte}(a), no
characteristic WGM features can be observed for TM-polarized
emission in the PL spectrum of a 135~nm-thick microdisk. For the TE
polarization the simulations predict the existence of two radial
families (radial mode numbers $p$=1,2). However, the measured
TE-spectra (Fig.\ref{foto} and Fig. \ref{tmte}(a)) show only the
$p$=1 modes, because the second family is much less intense. Thus,
all the observed spectral peaks belong to the same family and their
corresponding azimuthal mode numbers extend from $m$=29 (928~nm) to
$m$=42 (710.5~nm), with an average mode spacing of $\sim$15~nm.

We have prepared another series of microdisk arrays with a core
thickness of $d$=300~nm in order to check the existence of
TM-polarized WGM emission in thicker samples. In this case, slab
calculations give $n_{eff}$(TM)=1.45 (with $n_{eff}$(TE)=1.6),
suggesting a sufficient mode confinement for TM-polarized light.
Figure \ref{tmte}(b) shows both TE- and TM-polarized (inset) PL
spectra from an individual microdisk. As it was expected, now we
clearly observe the fine features of WGM in the TM-polarization. On
the other hand, higher order family modes with $p\geq$2 can be
resolved in the TE-spectrum.

\section{Pump power induced mode broadening: The role of Excited Carrier Absorption}
\begin{figure}
\centering\includegraphics[width=8cm]{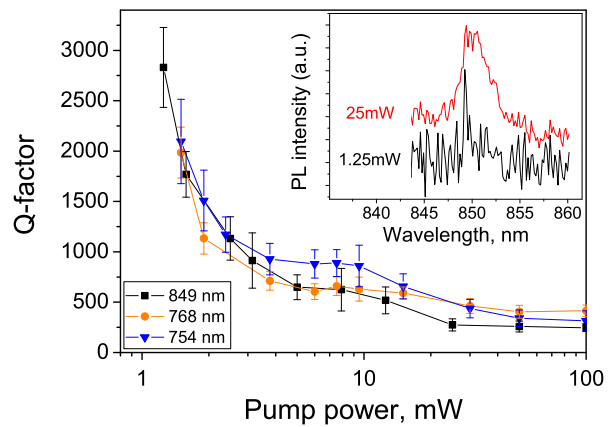} \caption{ (color
online) The measured $Q$-factors at increasing pump power are
plotted at three different wavelengths, reporting an order of
magnitude variation between two extreme pump powers. The inset shows
the WGM mode at $\lambda=849$~nm at the lowest and at a high pump
powers.} \label{power}
\end{figure}

Finally, we address some issues related to the influence of pumping
power on the WGM characteristics of our microdisks, in particular,
the observed significant linewidth modification. Figure \ref{power}
reports the measured $Q$'s of few distinct resonances at
$\lambda$=754~nm, 768~nm and 849~nm ($m=$39, 38 and 33,
respectively) of thin microdisks. We observe a monotonic ($\approx$
13-fold) decrease of $Q$ factors as the pump power increases from
1.25 to 100~mW. Such an impressive result needs a special attention.

The attenuating $Q$ factors suggest that at higher excitation powers
we either introduce an additional loss source or enhance the
existing ones. In a microdisk resonator, the total loss, resulting
from different loss mechanisms, is expressed through the sum of
inverse of possible limiting $Q$ factors:
\begin{equation}\label{qu}
    Q^{-1}=Q_{rad}^{-1}+Q_{mat}^{-1}+Q_{ssc}^{-1}+Q_{sa}^{-1},
\end{equation}
where the inverse of $Q_{rad}$, $Q_{mat}$, $Q_{ssc}$ and $Q_{sa}$
corresponds to radiation, material (bulk absorption and
propagation), surface scattering and surface absorption losses,
respectively. In our case, the last two terms can be considered
independent on the pump power within a good approximation. Thus, we
focus on the remaining terms $Q_{rad}$ and $Q_{mat}$ which can be
possibly modified by the excitation power. $Q_{rad}$ is related to
the disk geometry, while the second one is calculated as
$Q_{mat}=\frac{2\pi n_{eff}}{\lambda\alpha}$, with $\alpha$ being
the material loss coefficient.

First of all, from the comparison of WGM spectra measured at lowest
and high pump powers no relative spectral shift of resonances
neither modification of mode spacing is observed. Therefore, we
exclude the possibility of changes in the effective indices or the
disk sizes due to thermal heating effects. Thus, the power-dependent
$Q$ can arise from variations in $\alpha$ within the
nanocrystal-based disk material.

In recent studies on nc-Si containing waveguide systems, pump power
induced losses at 1.5$\mu$m wavelength have been addressed and
attributed to excited carrier absorption (ECA) \cite{dani,elliman}.
We believe that the same mechanism occurs here; Si nanocrystals
absorb strongly at visible wavelengths creating an exciton. Part of
the excitons can successively absorb other photons to promote
electrons to higher energetic levels in the nanocrystal conduction
band. Such absorption events will enhance the cavity losses, causing
the observed WGM broadening.

While under low pump conditions the ECA loss is expected to increase
linearly with power, $P$, at high pump powers different phenomena
can deviate this simple relationship; a number of processes, such as
the ECA itself (re-absorption of either a pump or emitted photon by
an already formed exciton), Auger recombination and the saturation
of the number of excitable nanocrytals will induce additional
non-linear $N(P)$ behavior. This non-linearity is clearly observable
in all our experimental data (Fig. \ref{power}).

Assuming a power-dependent loss, $\alpha=\alpha_{0}+\alpha^*(P)$,
one can rewrite Eq. \ref{qu} as
\begin{equation}\label{exp}
    Q_{exp}^{-1}(P)=\left( \sum Q_{rad,ss,sa}^{-1}+\frac{\lambda \alpha_0}{2\pi n_{eff}}\right)+\frac{\lambda \alpha^*(P)}{2\pi n_{eff}}.
\end{equation}

\begin{figure}
\centering\includegraphics[width=7cm]{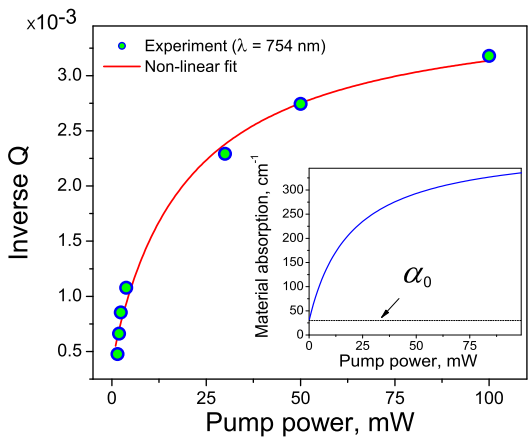} \caption{ (color
online) The non-linear fit of the measured inverse $Q_{exp}$ at
$\lambda=754~$nm ($m=$39) using Eq. \ref{exp}. The inset shows the
corresponding power-dependent absorption coefficient
$\alpha=\alpha_{0}+\alpha^*(P)$.} \label{fitQ}
\end{figure}

Here $\alpha_0$ stands for the material passive loss coefficient,
which has been independently estimated from ellipsometric data and
at $\lambda=754~$nm is of the order of 30~cm$^{-1}$ or less
\cite{note3}. The first two terms on the right part of Eq. \ref{exp}
represent the inverse of Q of the passive microdisk and, as
mentioned above, can be considered as a constant term. With this,
for the measured data $Q_{exp}^{-1}(P)$ a non-linear fitting
function in the form $\alpha^*(P)\sim aP(1+bP)^{-1}$, based on the
rate equation model \cite{elliman}, has been applied (Fig.
\ref{fitQ}).

The quality of the fit is further confirmed through the following
procedure. We consider the fitting value for the passive $Q$ (the
constant term in Eq. \ref{exp}), the FDTD-simulated geometrical
quality factor $Q_{rad}\approx1.1\times10^4$ for 8~$\mu$m disks and
negligible surface scattering/surface absorption losses, and
back-calculate the material passive loss coefficient. Such obtained
$\alpha_0\approx$32~cm$^{-1}$ is in very good agreement with the
ellipsometric results.

The situation gets more complex when due to the cavity effect the
spontaneous emission signal gets strong enough to affect the exciton
population, as in the case of stimulated emission. When this becomes
the dominant mechanism influencing the system loss, one expects that
the absorption grows sublinearly with power (increasing
transparency), leading to an inversion in the tendency of the $Q(P)$
curve (mode narrowing at high powers); hence, it would be possible
to achieve net gain and eventual lasing at higher pump powers.

Thus, the possible presence of different loss mechanisms result in
the observed complex behavior of $Q(P)$. In particular, we observe
from Fig.\ref{power}, that all $Q(P)$ curves show a clear
``shoulder" at increasing pump powers from $4$ to $12~$mW. A
confident model and detailed studies for quantification of these
phenomena are currently in course.

We stress that ECA will figure as the main limiting factor for a
possible multiwavelength lasing from the nc-Si-based microdisk. The
SRO material optimization (low-loss, positive material gain) should
play a key role for further enhancement of the observed $Q$-factors
of few thousands. Even with low, while inhomogeneously broadened
gain spectrum of nc-Si, microdisk resonators with similar $Q$'s
should be potential candidates to allow for a low-threshold laser
action, in a similar way as in III-V semiconductor micro-disk/pillar
devices \cite{lasers,pillar}.

\section{Conclusions}
To conclude, we reported on PL emission properties of
\emph{individual}, optically active microdisk resonators with Si
nanocrystals.  We observed subnanometer whispering-gallery
resonances in visible light emission with quality factors in excess
of 2800 from single microdisk resonators. Moreover, the influence of
pumping power on the WGM narrowing has been addressed, showing more
than an order of magnitude enhancement of Q-factors due to an
attenuation of pump-induced loss mechanisms. Both qualitative and
quantitative analysis suggest that the excited carrier absorption
stands behind the observed phenomenon.

\section*{Acknowledgments}
We acknowledge A. Picciotto and M. Wang for the help with PECVD
deposition of wafers and L. Vanzetti for helpful discussions. This
work has been partially supported by EC through the projects
PHOLOGIC (FP6-017158) and LANCER (FP6-033574), and INTEL.

\end{document}